\magnification 1200
Sznajd sociophysics model on a triangular lattice: ferro and antiferromagnetic 
opinions
\bigskip
Iksoo Chang

\bigskip
Department of Physics, Pusan National 
University

Pusan 609-735, Korea

chang@random.phys.pusan.ac.kr

\bigskip
\noindent
The Sznajd sociophysics model is generalized on the triangular lattice
with pure antiferromagnetic opinion and also with both ferromagnetic
and antiferromagnetic opinions. The slogan of the trade union
"united we stand, divided we fall" can be realized via the propagation
of ferromagnetic opinion of adjacent people in the union, but the
propagation of antiferromagnetic opinion can be observed among the
third countries between two big super powers or among the family
members of conflicting parents. Fixed points are found in both
models. The distributions of relaxation time of the mixed model are disperse and become
closer to log-normal as the initial concentration of down spins approaches 0.5,
whereas for pure antiferromagnetic spins 
they are collapsed into one master curve which is roughly lognormal.
We do not see the phase transition in the model.
\bigskip

The slogan of trade union "united we stand, divided we fall" was quantified in 
the Sznajd model[1] by observing the propagation for ordering of spins on
a lattice. Each site of a one- or two- dimensional lattice carries a spin which
can be either up or down and represents one of two possible opinions. If two
adjacent spins on the lattice have the same sign, the adjacent people have the
same opinon (ferromagnetic opinion) and convince the neighboring people to have
the same opinion. Therefore, the neighboring spins are forced to have the same sign.
If two adjacent spins are in the opposite direction (antiferromagnetic opinion),
the neighbor of left(right) spins take the value of right(left) spin of a given pair
of spins. The former may describe the way how the trade union unites in order
to achieve the better treatment, whereas the latter model may describe the union
with some members who always object to the common resolution or the small
countries between the Soviet Union and USA in the period of cold war. 
Several modifications of the Sznajd model were studied for the square lattice
[2,3,4]. 

A realization of the Sznajd model on a triangular lattice with the antiferromagnetic
opinion may represent some interesting social groups since the antiferromagnetic
spin on this lattice has the frustration effect with which not all neighbor pairs
can be antiparallel. 
A pair of two spins on a triangular
lattice has 8 neighbor spins. In the case of spreading of ferromagnetic opinions,
we do not expect anything special different from that on a square lattice;
Stauffer found a phase transition for two of the rules discussed in
ref [2].
The spreading of antiferromagnetic opinion, however, on a triangular lattice may
describe the groups who do not know what to do between two super powers or
between Republican and Democrat since they are frequently influenced so that they do
not know what to decide. 6 neighbor spins of the left(right) spin of a given 
horizontal pair of
two spins take the value of the right(left) spin. Therefore, after two adjacent
(antiferromagnetic) people convince 12 neighbors, 2 spins out of 12 neighbor spins
take a random value depending on the order of convincing. Therefore, these 2
spins do not know what to do, and they only depend on the party which influenced last. 
The spreading of pure antiferromagnetic opinion can be studied by
applying the Sznajd model on a triangular lattice, and also the spreading of mixed
(ferro + antiferro) opinion will be interesting to look at. 

We start with half of spins up and half of spins down on a triangular lattice. 
If one changes the initial concentration $p$ of down spins, one may study whether
there are fixed points for each $p$ and also the distribution of relaxation time after which
everybody shares the same opinion which can be ferromagnetic and antiferromagnetic
one. We averaged over $10^3$ samples of size $L \times L$ with $L=61$ where at 
each time
step a pair of two spins is chosen, and 8 neighbor spins are forced to take the
same or opposite sign depending on the spreading of ferromagnetic or antiferromagnetic
opinion. We first study the effect of antiferromagnetic spreading to that of
ferromagnetic spreading by employing both interactions between two spins.
Namely, when two adjacent spins are parallel, they spread a ferromagnetic opinion.
Otherwise, an antiferromagnetic opinion is spreading.  
We found this mixed model to reach always a fixed point at which the spins are all up or
down which means that the ferromagnetic opinion spreads throughout the lattice
(or trade union). Thus, the effect of antiferromagnetic opinion in the trade union
is irrelevant as long as there exists a ferromagnetic opinion in there. This reflects
that although there are some people who always object to 
the decision of the trade union,
the trade union will eventually achieve the final goal of unanimity. Figure
1 shows the distribution of relaxation times for the different initial concentration
of down spins on triangular lattices. The distributions for $p > 0.5$ is the same
as those for $p<0.5$, and the distributions become more log-normal as $p$ approaches
to $0.5$ and deviate gradually from log-normal distribution for $p \rightarrow 0$ or
$p \rightarrow 1$. 

When we allow the spreading of antiferromagnetic opinion only, 
the neighbor spins of left(right) spin take the
value of right(left) spin only for a given two adjacent opposite spins while 
they do not change their values for a given two adjacent parallel spins
which is different from the usual antiferromagnetic interaction of
two spins in Ising model.  
When the antiferromagnetic opinion propagates throughout a lattice (trade union), 
the spins on a triangular lattice and also on each of three sublattices  
order antiferromagnetically up and down sequentially for a given line.
Therefore, for a helical boundary condition and the system size $L=61$
the total staggered-magnetization at and after the fixed point, when an 
antiferromagnetic opinion spreads
completely, becomes either $L^2$ or $-L^2$. The relaxation time is measured as the
time for reaching the perfect antiferromagnetic ordering in the system. For $L=62$
there is no fixed point. Figure 2 shows the distribution of relaxation time for
antiferromagnetic case as we change the initial concentration $p$ of down spins. 
It is interesting to note that all the curves seem to collapse into one curve
which is roughly log-normal. Therefore, a pure antiferromagnetic model does not
care about the initical condition. The social group whose two adjacent members
do not share the same opinion cannot and will not reach any kind of the common
goal which everybody can agree, and so it takes a 
long time to crash their organization. 
And, it does not matter how high the concentration of initial down spins is.

In summary, we studied the generalization of Sznajd model to a triangular lattice
with spreading of mixed opinion and with the pure antiferromagnetic opinion.
For the mixed case of ferromagnetic and antiferromagnetic opinion, we always found
the fixed points for all values of the initial 
concentration of down spins. The effect of
antiferromagnetic opinion in the trade union, mixed with ferromagnetic
opinion, is irrelevant. Therefore, the trade union should be
generous enough to allow some people to object to the decision of trade union.
Otherwise, this trade union can not be called a democratic one. 
For the pure antiferromagnetic case, there are always fixed points, but this is
for the spreading of the opposing opinions. The distribution of relaxation time
is log-normal, but the relaxation time is much longer which means that it takes
a long time to destroy a common opinion. 
For both models we do not find the phase
transition, similar to the one-dimensional case of the original
Sznajd model[1].

\smallskip
The author is grateful to D. Stauffer for drawing his attention to 
the Sznajd sociophysics model, and acknowledges a fruitful discussion.

\bigskip
\parindent 0pt
[1] K. Szanajd-Werson and J. Sznajd, {\it Int. J. Mod. Phys. C} 
{\bf 11}, 1157 (2000)

[2] D. Stauffer, A.O. Sousa and S. Moss de Oliveira, {\it Int. J. Mod. Phys. C}
{\bf 11}, 1239 (2001)

[3] A.A. Moreira, J.S. Andrade Jr. and D. Stauffer, {\it Int. J. Mod. Phys. C}
{\bf 12}, 39 (2001) 

[4] A.T. Bernardes, U.M.S. Costa, A.D. Araujo and D. Stauffer, 
{\it Int. J. Mod. Phys. C} {\bf 12}, 159 (2001)

\bigskip
Captions:

Fig.1: The distribution of relaxation time for the mixed model with the
system size $L=61$. For $p$ close
to $0.5$ it is log-normal whereas it deviates as $p$ moves away from $0.5$. 
$p=0.05, 0.1, 0.2, 0.3, 0.5$ from the left to right curve.
For concentration $1-p$ we get the same results. 

Fig.2: The distribution of relaxation time for the pure antiferromagnetic model
with the system size $L=61$.
For all values of $p=0.05$ to $0.95$ the distribution collapses into one curve,
and the relaxation is much longer compared to those of the mixed model.

\end